\documentclass[aps,a4paper]{revtex4}

\usepackage{epsfig}
\usepackage{graphicx}
\usepackage{amsmath,amssymb,color}
\usepackage[english]{babel}

\parskip=\medskipamount

\newcommand{\eq}[1]{(\ref{#1})}
\newcommand{\fig}[1]{Fig.\ref{#1}}

\newcommand{\be}{\begin{equation}}
\newcommand{\ee}{\end{equation}}

\newcommand{\barr}{\begin{array}}
\newcommand{\earr}{\end{array}}

\newcommand{\beqn}{\begin{eqnarray}}
\newcommand{\eeqn}{\end{eqnarray}}

\newcommand{\bs}{\begin{subequations}}
\newcommand{\es}{\end{subequations}}

\newcommand{\bw}{\begin{widetext}}
\newcommand{\ew}{\end{widetext}}

\begin{document}

\title{Polygon-based hierarchical planar networks based on generalized Apollonian construction}

\author{M.V. Tamm$^{1,2}$, D.G. Koval$^1$, V.I. Stadnichuk$^1$}

\affiliation{$^1$ Faculty of Physics, Moscow State University, 119992, Moscow, Russia \\$^2$ Department of Applied Mathematics, MIEM, National Research University Higher School of Economics, 123458, Moscow, Russia
}

\date{\today}

\begin{abstract}

Experimentally observed complex networks are often scale-free, small-world and have unexpectedly large number of small cycles. Apollonian network is one notable example of a model network respecting simultaneously having all three of these properties. This network is constructed by a deterministic procedure of consequentially splitting a triangle into smaller and smaller triangles. Here we present a similar construction based on consequential splitting of tetragons and other polygons with even number of edges. The suggested procedure is stochastic and results in the ensemble of planar scale-free graphs, in the limit of large number of splittings the degree distribution of the graph converges to a true power law with exponent, which is smaller than 3 in the case of tetragons, and larger than 3 for polygons with larger number of edges. We show that it is possible to stochastically mix tetragon-based and hexagon-based constructions to obtain an ensemble of graphs with tunable exponent of degree distribution. Other possible planar generalizations of the Apollonian procedure are also briefly discussed.

\end{abstract}

\maketitle

\section{Intrduction}

It is often convenient to present big volumes of data as a graph, i.e., as a set of objects and binary relations (bonds) between them. This approach naturally arises in numerous contexts ranging from physics of disordered systems\cite{disorder_review} and biology\cite{sneppen_book} to sociology\cite{jackson_book} and linguistics(see, e.g., \cite{kenett,stella,valba}). The rapid growth in information technology ensures that larger and larger datasets of this type are becoming available. This naturally stimulates interest in the tools to analyse these datasets and simple (or not so simple) reference mathematical models, which can be used to probe their properties. Thus a rapid development in the last 20 years of a new interdisciplinary field on the boundary of random graph theory, data analysis and statistical physics, known as complex network theory\cite{newman, barabasi, dorogovtsev}. 

Among structural characteristics typical for many experimentally observed networks there are three especially common and striking (see, e.g.,\cite{newman}): (i) small world property (very small average node-to-node distance measured along the network), (ii) extremely wide, approximately power-law distribution of the node degrees (the networks with this property are often called `scale-free’), and (iii) large, as compared to referent randomized networks, concentration of short circles (e.g., triangles). It is reasonably easy to construct a model network that has one or two of these characteristics, e.g., Erdos-Renyi graphs\cite{er, krapivsky} are small world, random geometrical graphs\cite{bollobas} have large clustering coefficient, Barabasi-Albert model\cite{ba} generates small world scale-free networks, Watts-Strogatts model\cite{ws} --- small-world graphs with large clustering coefficient, etc. Generating all three properties simultaneously is much harder. Random geometric networks in hyperbolic space\cite{krioukov1, krioukov2, krioukov3} constitute one example of networks with these properties. Another one is the Apollonian network. 

\begin{figure}
\epsfig{file=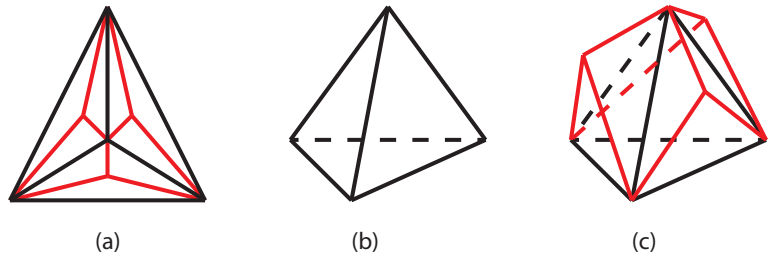, width=10 cm} \caption{Apollonian network. (A) First (black) and second (black and red) generations of the Apollonian network; (B) first generation Apollonian network is a tetragon; (C) Apollonian networks of higher generations can be thought of as trees constructed from adjacent teragons, here second generation is shown.}
\label{fig:1}
\end{figure}

The Apollonian network\cite{apol1, apol2} is a planar graph which arises naturally as a network representation of the Apollonian packing of the plane, see \fig{fig:1}. The construction of this network can be explained recursively as follows. Take a triangle and pick a point inside it, and connect it to the three corners of the triangle. As a result you get a set of 3 adjacent triangles which form the 1-st generation Apollonian network. Now, pick a point inside each of the four triangles, and connect it to the respective corners, this gives a 2-nd generation Apollonian network, then repeat {\it ad infinitum}. 
This network has been studied extensively in recent years and it has been shown to have many beautiful properties. For example, the degree distribution and clustering coefficient has been calculated\cite{apol1}, as well as the average path length\cite{path}. Notably, Apollonian network can be reinterpreted as a simplicial complex in a following way\cite{bianconi}. 1-st generation Apollonian network is a tetrahedron (3-simplex). In turn, 2-nd generation Apollonian network consists of 4 tetrahedra: the original one and another three, each having a common 2-face with the original one, 3-rd generation Apollonian network consists of 13 tetrahedra: 1 produced in the first generation, 3 produced to 2-nd generation and 9 new ones attached to each free face of those three. Thus, one can think of an Apollonian network as a regular rooted tree of tetrahedra touching each other by common faces (see \fig{fig:1}(b,c)).
Many properties of the Apollonian network can be calculated exactly, which makes it a nice toy model for the study of various properties of real scale-free networks. As a result, there have been a significant number of papers in recent years studying percolation \cite{bianconi}, spin models \cite{magnetic}, signal spreading \cite{gossip}, synchronization\cite{evolving}, traffic\cite{traffic}, random walks\cite{walks1,walks2} etc. on the Apollonian network.

Despite being such a beautiful and well-studied object, the Apollonian network has certain drawbacks as a model of real networks. Most importantly, it is a single deterministic object with certain fixed properties, e.g., a fixed degree distribution with a fixed power law exponent $\gamma = \ln 3/\ln 2$. Importantly, that degree distribution is actually not a true power law, but rather a log-periodic distribution consisting of a sequence of atoms at points $3 \times 2^n$ and a power-law envelope. This means that the network is scale-invariant only with respect to certain discrete renormalizations, and thus do not have the full set of properties of a true power law distribution (see \cite{newberry} for a recent discussion of the topic). One natural generalization is a random Apollonian network \cite{random_arch,random2,random3}, which is constructed, instead of a regular generation-by-generation process, by sequential partitioning of arbitrary chosen triangles. The average degree distribution in such network is a true power law with exponent $\gamma_R = 3$ \cite{random2}. Another way of generalizing the network is to consider as the building blocks of the network construction procedure the k-simplexes with $k>3$, which gives rise to multidimensional Apollonian networks \cite{multidim1, multidim2}.

In this paper we suggest another way of generalization of the Apollonian network construction and present family of random planar network models, all of them small-world and scale-free. The main idea is to construct an Apollonian-style iteration procedure based on polygons with different numbers of edges. The presentation is organized as follows. In section II we start with constructing a tetragon-based Apollonian-style network, and explicitly calculating its degree distribution. We then generalize the suggested procedure to polygons with arbitrary even number of edges. In section IV we show that it is possible to construct a continuous one-parametric family of models interpolating between the tetragon- and hexagon-based models, and show that the models in this family have a power low degree distribution with exponent depending on the parameter, so that it is possible to adjust it to fit the desired degree distribution (note that adjustable exponent of the degree distribution can be obtained by different means in the so-called Evolving Apollonian networks \cite{evolving, random3}). In the discussion section we summarize the results of the paper, and discuss further possible generalizations.

\section{Tetragon-based network}

\subsection{Definition}

%Instead of making a straightforward generalization (namely, putting different number of additional
%vertices on new edges) we suggest here one which is, actually, a significant variation of the main idea in a
%sense that it has a substantial random element.

\begin{figure}
\epsfig{file=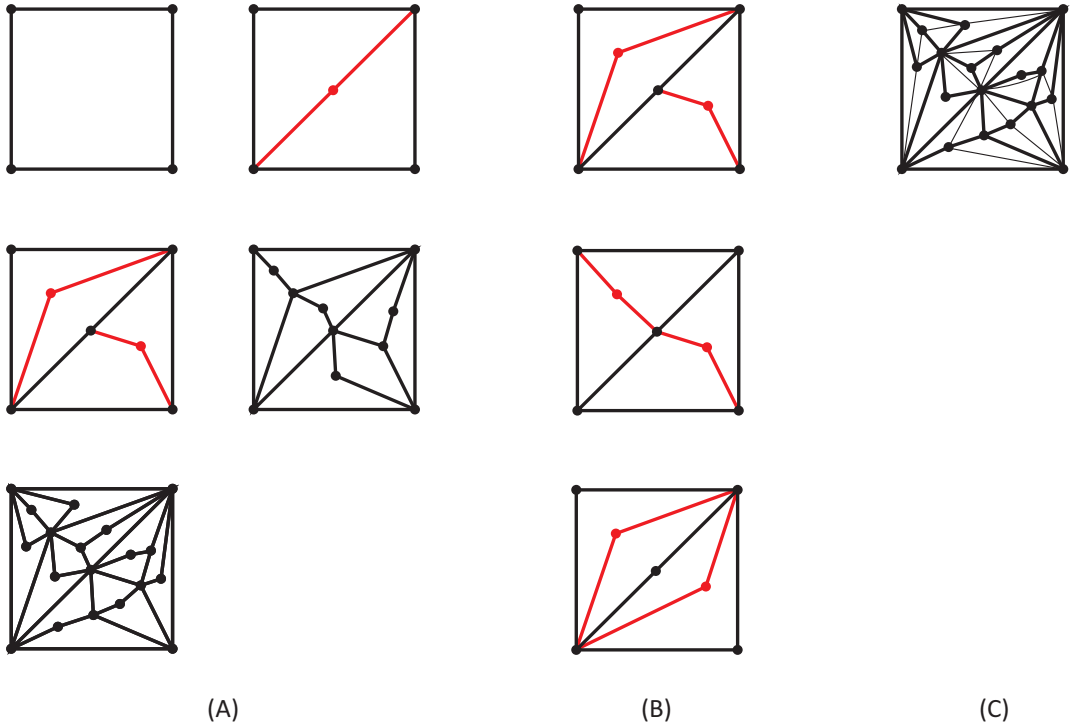, width=14 cm} \caption{Construction of a tetragon-based network. (A) Representatives of the tetragon-networks up to the 4th generation; (B) three possible topologically different realizations of the second-generation network; (C) random triangulation of the 4th generation network.}
\label{fig:3}
\end{figure}

Among several possible ways of generalizing the procedure described above to the case of polygons, consider the following. Take a tetragon and pick a point inside it; then choose
(at random) a pair of non-adjacent vertices of the original tetragon and connect them with the new point inside. We now have two adjacent tetragons, for which we can repeat this construction as shown in 
\fig{fig:3}. Importantly, contrary to the standard Apollonian network, which is a deterministic object, the network resulting from this procedure is stochastic. Indeed, already in the second generation there are three topologically different realizations of the network, see \fig{fig:3}(B). Notably, at any generation this network has no triangles, and is, in fact, bipartite. 

\subsection{Degree distribution}

The natural question to ask about this newly introduced class of planar Apollonian-like networks, is what is the degree distribution of the nodes $G_n(k)$ in the $n$-th generation of the network, and what distribution ${\cal G}(k)$ it converges to for $n \to \infty$. By analogy with the Apollonian networks one expects ${\cal G}(k)$ to be scale free, i.e.
\be
{\cal G}(k) \simeq C k^{-\alpha}\;\;\; k\gg 1
\label{scalefree}
\ee
with some yet unknown constants $C$ and $\alpha$. 

To calculate the degree distribution $G_n(k)$ note that the degree of any given node is a random variable, whose distribution $F_{n-m}(k)$, for all nodes except the four initial ones, depends only on the number of generations between the generation $m$ at which it was created, and current generation $n$. Indeed, each node with degree $k$ has exactly $k$ adjacent tetragons ($k-1$ for the four initial nodes), and at every step of the recurrent procedure each of these tetragons is split in two, which results in a creation of a new edge adjacent to the node with probability 1/2 (in the other half of the case the splitting path does not go through the given node). These splitting events happen independently for all tetragons. The overall degree distribution is therefore calculated by averaging over degree distributions of different generations: 
\be
G_n^{all}(k) = \dfrac{4F^{(0)}_{n}(k) + \sum_{m=1}^n Q_m F_{n-m}(k)}{4+ \sum_{m=1}^n Q_m} =\frac{4}{2^n+3}F^{(0)}_{n}(k) +\frac{2^n-1}{2^n+3}G_n(k),\;\; G_n(k)= \frac{\sum_{m=1}^n Q_m F_{n-m}(k)}{\sum_{m=1}^n Q_m}
\label{gf}
\ee
where $Q_m = 2^{m-1}$ is the number of nodes created in the $m$-th generation, and $F^{(0)}_{n}(k)$ is the degree distribution of the four initial nodes, and it is convenient to introduce $G_n(k)$, the degree distribution of all nodes except four initial ones. 

\subsection{Recurrence relation for $F_n(k)$}. 

It is easy to construct the recurrence relation for $F_n(k)$. Indeed, let $l$ be the degree of a node in the $(n-1)$-th generation. This means that this node has $l$ tetragons adjacent to it, and when constructing the $n$-th generation of the network each of them will be split in half, and with probability 1/2 the splitting path will go through the node under consideration. Every such path increases the degree of the node by one. Thus, the overall degree may increase by $l'$, $0\leq l' \leq l$, with probability $2^{-l} {l\choose l'}$, leading to
\begin{equation}
F_{n}(k) = \sum \limits_{l=[(k+1)/2]}^k 2^{-l} {l \choose {k-l}} F_{n-1}(l)\text{ for } n\geq 1;\;\;
F_0(k) = \delta_{k,2};
\label{recurrence_f4}
\end{equation}
\\
where we allowed for the fact that all nodes are created with degree 2.

Introduce a generating function
\begin{equation}
f_{n}(\lambda) = \sum\limits_{k=2}^{\infty} \lambda^{k} F_{n}(k).
\end{equation}\\
Then, substituting \eq{recurrence_f4}, one gets $f_0(\lambda) = \lambda^2$ and
\begin{equation}
\begin{array}{rll}
f_{n}(\lambda) &= & \displaystyle \sum\limits_{k=2}^{\infty} \sum \limits_{l=[(k+1)/2]}^k \lambda^k 2^{-l} {l \choose {k-l}} F_{n-1}(l) = \medskip \\
&=& \displaystyle \sum\limits_{l=2}^{\infty} \sum\limits_{m=0}^{l} (\lambda/2)^{l} {l \choose {m}} \lambda^{m} F_{n-1}(l) = \sum\limits_{l=2}^{\infty} \left(\frac{\lambda(1+\lambda)}{2}\right)^{l} F_{n-1}(l)=f_{n-1}\left(\frac{\lambda(1+\lambda)}{2}\right),
\end{array}
\label{f_rec}
\end{equation}
where we changed the order of summation, introduced $m=k-l$, and used the binomial formula 
\be
(1+\lambda)^l = \sum_{m=0}^l {l\choose m} \lambda^m.
\ee
The recurrence relation for the four initial nodes is a bit different, because in their case the node of degree $l$ has only $l-1$ adjacent tetragons:
\begin{equation}
F_{n}^{(0)}(k) = \sum \limits_{l=[k/2]+1}^k 2^{-l+1} {l-1 \choose {k-l+1}} F_{n-1}^{(0)}(l)\text{ for } n\geq 1;\;\;
F_0(k) = \delta_{k,2};
\label{recurrence_f4_0}
\end{equation}
It is easy to get the following equation for the generating function
\be
f^{(0)}_{n}(\lambda) =\sum\limits_{k=2}^{\infty} \lambda^{k} F_{n}^{(0)}(k)= \frac{2}{1+\lambda} f^{(0)}_{n-1}\left(\frac{\lambda(1+\lambda)}{2}\right)
\label{f0_rec}
\ee
In the $n \to \infty$ limit both $f_{n}(\lambda)$ and $f^{(0)}_{n}(\lambda)$ converge to zero for all $|\lambda|<1$. Indeed, the probability to have any finite degree many generations after the creation of a node is exponentially small. 

\subsection{Generating function of the degree distribution}

Introduce know the generating functions, 
\be
g_n(\lambda) = \sum_k G_n(k) \lambda^k,\;\; g_n^{all}(\lambda) = \sum_k G_n^{all}(k) \lambda^k 
\ee
where $G_n(k), G_n^{all}(k)$ are given by \eq{gf}. The former of them satisfies recurrence relation 
\be
(2^{n+1}-1)g_{n+1}(\lambda) = 2^n\lambda^2 + (2^n-1)g_n\left(\frac{\lambda(1+\lambda)}{2}\right) \text{ for } n\geq 0;\;\; g_0(\lambda) = \lambda^2,
\label{g_rec}
\ee
%\be
%(2^n +3) g_n(\lambda) = 4f^{(0)}_{n}(\lambda) +\sum_{m=1}^n 2^{m-1} f_{n-m}(\lambda),
%\label{gf}
%\ee
which in the limit of large $n$ reduces to
\be
g_{n+1}(\lambda) = \frac{1}{2}\lambda^2 + \frac{1}{2} g_n\left(\frac{\lambda(1+\lambda)}{2}\right).
\label{g_rec2}
\ee
Contrary to \eq{f_rec} and\eq{f0_rec}, \eq{g_rec} has a non-trivial limiting solution for $n \to \infty$, which is the solution of a functional equation
\be
2 \bar g(\lambda) = \lambda^2 +{\bar g}\left(\frac{\lambda(1+\lambda)}{2}\right); \;\; {\bar g}(\lambda) = \sum_k {\cal G}(k) \lambda^k.
\label{g_rec2}
\ee
It seems impossible to solve this equation for all $\lambda$, but it is easy to study its behavior in the vicinity of $\lambda = 0$, and $\lambda = 1$, which control the small-$k$ and large-$k$ behavior of ${\cal G}(k)$, respectively. For small $\lambda$, substituting
\be
\bar g(\lambda) = p_2 \lambda^2 + p_3 \lambda^2 + p_4 \lambda^4 +\dots 
\label{g_l0}
\ee
into\eq{g_rec2} one gets
\be
p_2 = \frac{4}{7}; \; p_3=\frac{16}{105};\; p_4 = \frac{16}{155}; p_5 =\frac{64}{1519};\dots
\ee
for the limiting probabilities of having a node degree equal to 2, 3, 4, 5, $\dots$.

In turn, if the large-$k$ behavior of ${\cal G} (k)$ is given by \eq{scalefree}, then $\bar g(\lambda)$ is divergent at $\lambda=1$. Subtracting the divergent part of the function one gets
\be
\bar g(\lambda) = g_{reg}(\lambda) + C \text{Li}_{\alpha}(\lambda).
\ee
Expanding each term separately in powers of $\epsilon = 1- \lambda$ one gets
\be
\bar g(\lambda) = \sum_{i=0}^{\infty} a_i \epsilon^i + C \Gamma(1-\alpha) \epsilon^{\alpha-1} \sum_{i=0}^{\infty} b_i \epsilon^i,
\label{g_l1_expansion}
\ee
where values of $a_i$ are non-universal, and depend on the small-$k$ behavior of ${\cal G}(k)$, in particular $a_0=\sum_k {\cal G}(k) =1$, while $b_i$ are the coefficients of the expansion of the polylogairthm in the vicinity of 1:
\be
b_0 = 1,\;\; b_1 = \frac{1}{2}(\alpha - 1),\;\; b_3 =\frac{1}{24}(\alpha - 1) (3\alpha + 2 ),\dots
\label{polylog_expansion}
\ee 
Now substituting $\lambda(\lambda+1)/2 = 1 - 3 \epsilon/2 + \epsilon^2/2$ into \eq{g_rec2} and expanding in the vicintity of $\epsilon = 0$ we see that the two sums in \eq{g_l1_expansion} do not mix up, and the corresponding coefficients should be equated separately, leading to
\be
\begin{array}{rll}
2a_0 &=& 1 + a_0,\;\; a_0 = 1 \medskip \\
2a_1 &=& -2 + 3a_1/2 ,\;\; a_1 = -4 \medskip \\ 
2a_2 &=& 1 - a_1/2 + 9a_2/4, \;\; a_2 = -12 \medskip \\
\dots & & \dots \medskip \\
2 C \Gamma(1-\alpha) b_0 &= &C \Gamma(1-\alpha) (3/2)^{\alpha-1} b_0, \;\; \alpha=1+ \frac{\ln 2}{\ln(3/2)} = \frac{\ln 3}{\ln 3 - \ln 2} \approx 2.70951\dots
\end{array}
\ee
Now, since, $2<\alpha<3$, the 0-th and 1-st moments of the distribution converge, and are related to the first two coefficients in the expansion of $\bar{g}(\lambda)$:
\be
\sum_k {\cal G}(k) =a_0 = 1;\; \; \sum_k k {\cal G}(k) =\langle k\rangle_{\infty} = -a_1 = 4,
\ee 
while all the higher moments, starting from $\langle k^2\rangle$ diverge with growing $n$. 

It is instructive to calculate the exact values of moments $\langle k\rangle_n, \langle k^2\rangle_n$ for all finite $n$. To do this, note that 
\be
\langle k\rangle_n = \left. \frac{d g_{n} (\lambda)}{d\lambda}\right|_{\lambda=1};\;\; \langle k^2\rangle_n = \langle k\rangle_n + \left. \frac{d^2 g_{n} (\lambda)}{d\lambda^2}\right|_{\lambda=1}.
\label{moments}
\ee
Equation \eq{g_rec} implies 
\begin{equation}
(2^{n+1}-1) g_{n+1}^\prime(\lambda) = 2^{n+1}\lambda + (2^n-1)\left(\frac{2\lambda+1}{2}\right)
g_n^\prime\left(\frac{\lambda(1+\lambda)}{2}\right)
\end{equation}
which leads for $\lambda = 1$ to
\begin{equation}
\left(1- 2^{-n-1}\right) \langle k\rangle_{n+1} = 1 + \frac{3}{4} \left(1 - 2^{-n}\right) \langle k\rangle_n 
\end{equation}
Substituting
\be
b_n = 4 - \left(1- 2^{-n}\right) \langle k\rangle_n, 
\ee
and allowing for the initial condition $\langle k \rangle_1 = 2, b_1 = 3$, one gets
\be
b_{n} = \frac{3}{4}b_{n-1} = 4 \left(\frac{3}{4} \right)^n
\ee
and thus
\be
\langle k\rangle_n = 4 \frac{1 - (3/4)^n}{1-(1/2)^n}
\label{mean_k}
\ee
This is the average degree of all nodes except the original four at the $n$-th step of the network-generation process. Given \eq{gf} it is easy to calculate the average degree of {\it all} nodes: 
\be
\langle k\rangle_n^{all} = \sum k G_n^{all}(k) =\frac{(2^n-1)\langle k\rangle_n + 4(1+(3/2)^n)}{ 2^n+3}= 4\frac{(1 - (3/4)^n) 2^n + 1+(3/2)^n}{2^n+3}= 4\frac{2^n + 1}{2^n+3}.
\ee

\subsection{Second moment of the finite-generation distribution}

To calculate the second moment, take the second derivative of \eq{g_rec}
\be(2^{n+1}-1)g''_{n+1}(\lambda) = 2^{n+1}+(2^n-1)g'_n\left(\frac{\lambda(1+\lambda)}{2}\right) + (2^n-1)\left(\frac{2\lambda+1}{2}\right)^2 g''_n\left(\frac{\lambda(1+\lambda)}{2}\right)
\ee
and take into account \eq{moments}. Substituting $\lambda = 1$ and allowing for the fact that
\be
g'_n(1) = \langle k\rangle_n;\; g''_n(1) = \langle k^2\rangle_n -\langle k\rangle_n
\ee
leads to
\be
(2^{n+1}-1)(\langle k^2\rangle_{n+1} - \langle k\rangle_{n+1}) = 2^{n+1}+
(2^n-1)\langle k\rangle_n+\frac{9}{4}(2^n-1)(\langle k^2\rangle_n-\langle k\rangle_n) 
\ee
or
\be
(2^{n+1}-1)\langle k^2\rangle_{n+1} - \frac{9}{4}(2^n-1)\langle k^2\rangle_n = 2^{n+1} +(2^{n+1}-1)\langle k \rangle_{n+1} - \frac{5}{4}(2^{n}-1)\langle k \rangle_{n}, 
\ee
which, after substituting \eq{mean_k} simplifires to
\be
(2^{n+1}-1)\langle k^2\rangle_{n+1} - \frac{9}{4}(2^n-1)\langle k^2\rangle_n = 2^n\left( 5 - \left(\frac{3}{4}\right)^n \right).
\ee
Define now the sequence
\be
a_n = \langle k^2 \rangle_n \frac{2^n-1}{2^n}
\ee
and its generating function $F(s) = \sum a_{n} s^{n}$. The recurrency for $a_n$ reads
\be
a_{n+1} = \frac{9}{8}a_n + \frac{5}{2} - \frac{1}{2} \left(\frac{3}{4}\right)^n
\ee
for $n\geq 1$, and $a_1=2$. Then 
\be
F(s)\left(1-\frac{9}{8}s \right) = 2 s + \frac{5}{2} \frac{s^2}{1-s} - \frac{3}{8} \frac{s^2}{1- 3s / 4}
\ee
In order to proceed further, note that
\be
\begin{array}{rll}
\displaystyle \frac{2s}{1-9s/8} & = & \displaystyle -\frac{16}{9} +\frac{16}{9}\frac{1}{1-9 s/8} \medskip \\
\displaystyle \frac{s^2}{(1-s)(1-9s/8)} &=& \displaystyle \frac{8}{9} - 8 \frac{1}{1-s} +\frac{64}{9} \frac{1}{1-9s/8} \medskip \\
\displaystyle \frac{s^2}{(1-3s/4)(1-9s/8)} &=& \displaystyle \frac{32}{27} - \frac{32}{9} \frac{1}{1-3s/4} +\frac{64}{27} \frac{1}{1-9s/8}
\end{array}
\ee
Thus, 
\be
F(s) = \frac{56}{3}\frac{1}{1-9 s/8} - 20 \frac{1}{1-s} + \frac{4}{3} \frac{1}{1- 3s / 4}; \;\; a_n = \frac{56}{3} \left(\frac{9}{8}\right)^n - 20 + \left(\frac{3}{4}\right)^{n-1}
\ee
and
\be
\langle k^2\rangle = (1- 2^{-n})^{-1} \left[ \frac{56}{3} \left(\frac{9}{8}\right)^n - 20 + \left(\frac{3}{4}\right)^{n-1} \right ]
\label{k2}
\ee
Proceeding in the same way, one gets 
\be
\langle k^2\rangle^{(0)} = \frac{4}{3}\left(\frac{9}{4}\right)^n+\frac{5}{3}\left(\frac{3}{2}\right)^n+1
\ee
Thus, the total average degree is
\be
\begin{array}{rll}
\langle k^2\rangle_{all} &=& \displaystyle \frac{2^n-1}{2^n+3}\langle k^2\rangle +\frac{4}{2^n+3}\langle k^2\rangle^{(0)} =\medskip \\
& = & \displaystyle \left(1+\frac{3} {2^{n}}\right)^{-1} \left[ 24 \left(\frac{9}{8}\right)^n - 20 + 8 \left(\frac{3}{4}\right)^{n}+4 \left(\frac{1}{2}\right)^{n} \right ] \approx 24 \left(\frac{9}{8}\right)^n - 20,
\end{array}
\label{k2all}
\ee
where the approximal equality holds for $n \gg 1$. Comparing \eq{k2} and\eq{k2all} shows that, interestingly, the four initial nodes contribute a finite fraction to the overall value of $\langle k^2\rangle_{all}$, which converges to 2/9 for large $n$.

\subsection{Scaling form of the degree distribution}

Finally, one expects that for large $n$ the degree distribution has a limiting scaling form
\be
G_n(k) \approx C k^{-\alpha} \phi \left(\frac{k}{a^n}\right) = {\cal G}(k) \phi \left(\frac{k}{a^n}\right) \text{ for } k \gg 1
\ee
Here $a$ is easy to obtain from the large-$n$ behavior of $\langle k^2\rangle_n$
\be
\langle k^2\rangle_n = \sum k^2 G_n(k) \sim a^{n(3-\alpha)}, \;\; 
\ee
where we took into account that (contrary to the lower moments) $\langle k^2\rangle_n$ is controlled by the tail of the distribution. Substituting \eq{k2all} one gets
\be
a^{3-\alpha} = 9/8;\;\; a = 3/2,
\ee
which is to be expected since the typical maximal degree of the network increases by a factor $3/2$ on each step.

%\subsection {Simulations}

\begin{figure}
\epsfig{file=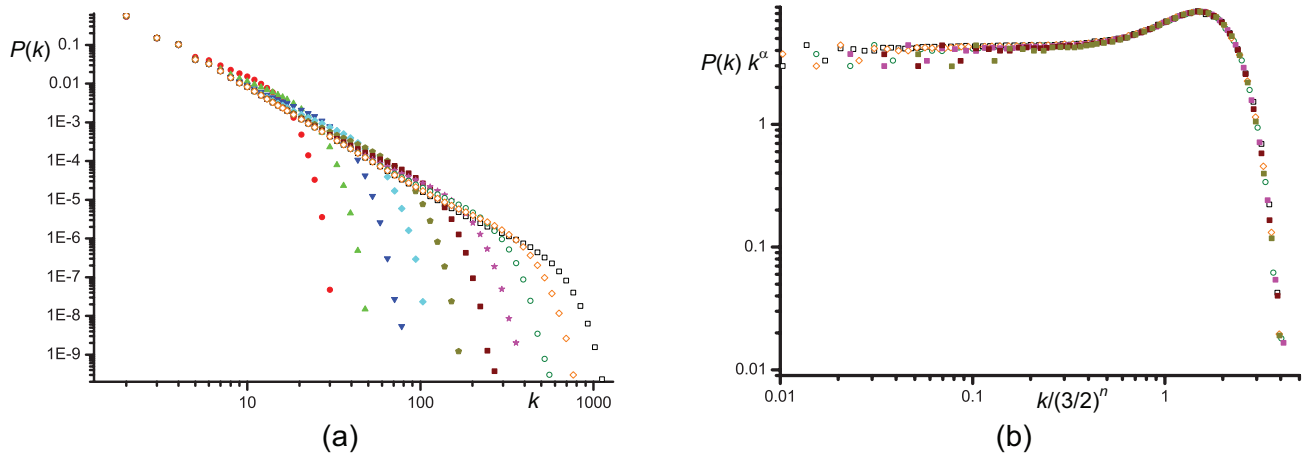, width=17 cm} \caption{(A) Degree distributions of various generations of the tetragon-based networks: generations 5 (red circles), 6 (green upward triangles), 7 (blue downward triangles), 8 (cyan diamonds), 9 (dark yellow pentagons), 10 (wine red squares), 11 (purple stars), 12 (green open circles), 13 (orange open diamonds), and 14 (black open squares). The results shown are after averaging over $2\times 10^5$ realizations and logarithmic binning with step 1.1. (B) Same distributions in the rescaled coordinates.}
\label{fig:degree4}
\end{figure}

In order to check the predictions of the model, we have generated $2\times 10^5$ realizations of the networks of up to 14-th generation. \fig{fig:degree4}(A) shows the resulting degree distributions for sequential generations of the network. It can be seen from \fig{fig:degree4}(B) that after renormalization of the axes by the factors $(3/2)^n$ and $k^{-\alpha}$ respectively, the data collapse perfectly on a single scaling curve $\phi(x)$.

%The network is small-world by the same arguments as before, and the scaling exponent $\gamma_4$ is once again
%easy to estimate. Indeed, the number of new vertices grows with the number of generation as $2^N$. In turn,
%the degree of the vertex is multiplied by 3/2 with each new generation (the degree of the vertex equal to the
%number of adjacent elementary tetragons, each of them should be partitioned in 2 parts in next generation,
%which will with probability 1/2 produce a new edge at the vertex under discussion). This immediately means
%that the degree distribution constant $\gamma_4 = \ln 2/ \ln (3/2) \approx 1.71$.

\section{Polygon-based networks for polygons with any even number of edges}

\begin{figure}
\epsfig{file=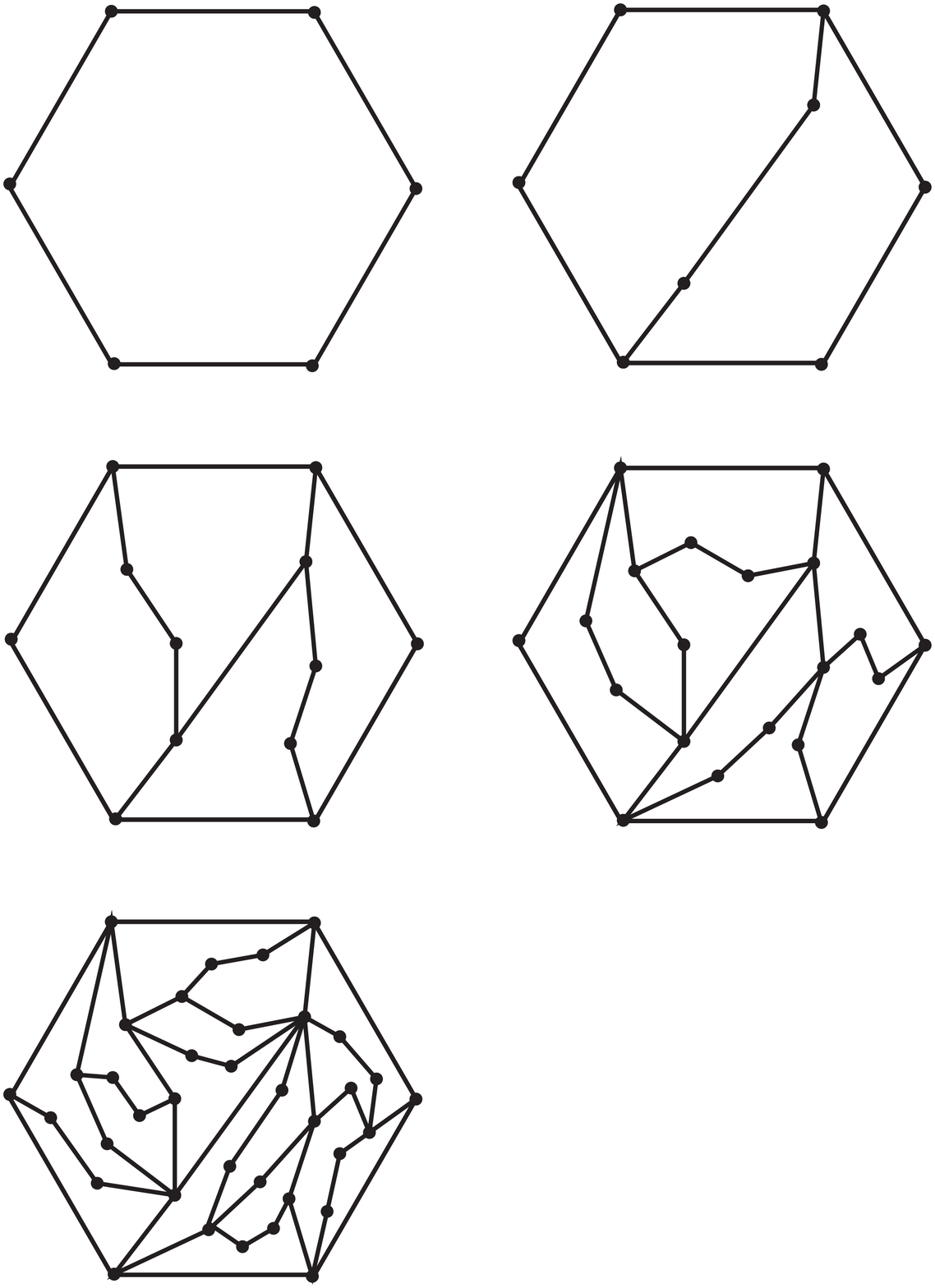, width=8 cm} \caption{Construction of the hexagon-based network (up to 4th generation).}
\label{fig:4}
\end{figure}

The procedure suggested in the previous section can be easily generalized for any even number of edges $2m$ ($m\geq 2$) in the generating polygon, see generalization for $m=3$ in figure \fig{fig:4}. We get thus a sequence of planar scale-free network models with degree distributions converging to 
\be
{\cal G}_m(k) \simeq C k^{-\alpha_m}\;\;\; k\gg 1
\label{scalefree_m}
\ee
with $m$-dependent exponents $\alpha_m$. At each generation each polygon is split by a path connecting directly opposite nodes. There are $m$ different ways of such a splitting, so each node of a polygon participates in the splitting with probability $1/m$. This allows to generalize the recurrence relations \eq{recurrence_f4} and \eq{recurrence_f4_0} for the degree distribution of a node $n$ generations after its creation in the following way
\begin{equation}
F_{n,m}^{(0)}(k) = \sum \limits_{l=[k/2]+1}^k {l-1 \choose {k-l+1}} \left(\frac{1}{m}\right)^{k-l+1} \left(\frac{m-1}{m}\right)^{2l-k-2} F_{n-1,m}^{(0)}(l)\;\;\;\;\; \text{ for } n\geq 1;\;\;
F_0^{(0)}(k) = \delta_{k,2};
\label{recurrence_fm}
\end{equation}
for the original $2m$ nodes, and
\begin{equation}
F_{n,m}(k) = \sum \limits_{l=[(k+1)/2]}^k {l \choose {k-l}} \left(\frac{1}{m}\right)^{k-l} \left(\frac{m-1}{m}\right)^{2l-k} F_{n-1,m}(l)\;\;\;\;\;\text{ for } n\geq 1;\;\;
F_{0,m}(k) = \delta_{k,2};
\label{recurrence_fm}
\end{equation}
for all the rest. The number of nodes created at $n$-th generation ($n \geq 1$) is $(m-1) 2^n$. Proceeding in exactly the same way as before, we get
\be
f^{(0)}_{n,m}(\lambda) =\sum\limits_{k=2}^{\infty} \lambda^{k} F_{n}^{(0)}(k)= \frac{m}{m-1+\lambda} f^{(0)}_{n-1}\left(\frac{\lambda(m-1+\lambda)}{m}\right),
\label{f0_rec_m}
\ee
\be
\begin{array}{rll}
g_{n,m}(\lambda) &=& \displaystyle \frac{1}{(m-1)(2^{n}-1)} \sum\limits_{l=1}^n \sum\limits_{k=2}^{\infty} (m-1)2^{l} F_{n-l,m}(k) \lambda^{k} = \medskip \\
&=& \displaystyle \frac{2^{n-1}}{2^n-1}\lambda^2 + \frac{2^{n-1}-1}{2^n-1} g_{n-1,m}\left(\frac{\lambda(m-1+\lambda)}{m}\right) \text{ for } n\geq 1;\;\; g_{1,m}(\lambda) = \lambda^2,
\end{array}
\label{g_rec_m}
\ee
\be
g_{n,m}^{all}(k) =\frac{2m}{(m-1)2^n+m+1}f^{(0)}_{n,m}(\lambda) +\frac{(m-1)(2^n-1)}{(m-1)2^n+m+1}g_{n,m}(\lambda)
\label{gf_m}
\ee
for the generating functions of the degree distributions of the original, newly created, and all nodes of the network, $f^{(0)}_{n,m}(\lambda)$, $g_{n,m}(\lambda)$, and $g_{n,m}^{all}(\lambda)$, respectively.
The limiting function
\be
\bar{g}_m(\lambda) = \lim \limits_{n\to\infty} g_{n,m}(\lambda)
\ee
satisfies
\begin{equation}
\bar{g}_m(\lambda) = \frac{\lambda^2}{2} + \frac{1}{2}\bar{g}_{m}\left(\frac{\lambda(m-1+\lambda)}{m}\right)
\label{limiting_m}
\end{equation}
and its behavior is easy to analyze both in the vicinity of $\lambda =0$ and $\lambda=1$. Expanding \eq{limiting_m} for small $\lambda$ one gets 
\be 
\begin{array}{rll}
p_2^m &= &\displaystyle \frac{m^2}{K_{2,m}};\;\; p_3^m =\frac{2m^3(m-1)}{K_{2,m} K_{3,m}};\;\; p_4^m = \frac{m^3(2m^3 +5(m-1)^3)}{\prod\limits_{l=2}^{4} K_{l,m}}; \medskip \\ 
p_5^m&=&\displaystyle \frac{m^5 (m-1)^2(12m^4 +8m^3(m-1)+14(m-1)^4)}{\prod\limits_{l=2}^{5} K_{l,m}};\;\;\dots 
\end{array}
\ee
where we introduced a short-hand notation $K_{l,m} =2m^l - (m-1)^l$.

In turn, in the vicinity of $\lambda=1$ $\bar{g}_m(\lambda)$ takes the form \eq{g_l1_expansion} with $b_i$ given by \eq{polylog_expansion}. Substituting the ansatz \eq{g_l1_expansion} into \eq{limiting_m} one gets
\be
\begin{array}{rll}
2a_0 &=& 1 + a_0,\;\; a_0 = 1 \medskip \\
2a_1 &=& -2 + (m+1)a_1/m ,\;\; a_1 = -2m/(m-1) \medskip \\ 
2a_2 &=& 1 - a_1/m + (m+1)^2 a_2/m^2, \;\; a_2 = \dfrac {(m+1)m^2}{(m-1)(m^2-2m-1)} \medskip \\
\dots & & \dots \medskip \\
2 C \Gamma(1-\alpha_m) b_0 &= &C \Gamma(1-\alpha_m) (m+1/m)^{\alpha_m-1} b_0, \;\;\alpha_m=1+ \dfrac{\ln 2}{\ln(m+1)-\ln m} 
\end{array}
\ee
Thus, for any $m \geq 3$ $\alpha_m >3$ and the second moment of ${\cal G}_m(k)$ converges. As a result, the moments are controlled by the coefficients $a_i$:
\be
\sum_k {\cal G}(k) =a_0 = 1;\; \; \sum_k k {\cal G}(k) = -a_1 = \frac{2m}{m-1}; \;\;\;\sum_k k^2 {\cal G}(k) =2a_2 - a_1 = \frac {2m(2m^2-m-1)}{(m-1)(m^2-2m-1)}. 
\label{moments}
\ee 
Since the second moment of ${\cal G}_m(k)$ is now controlled by the values of distribution at small $k$, the initial $2m$ nodes do not contribute to the second moment.

%\subsection{Simulations}

\begin{figure}
\epsfig{file=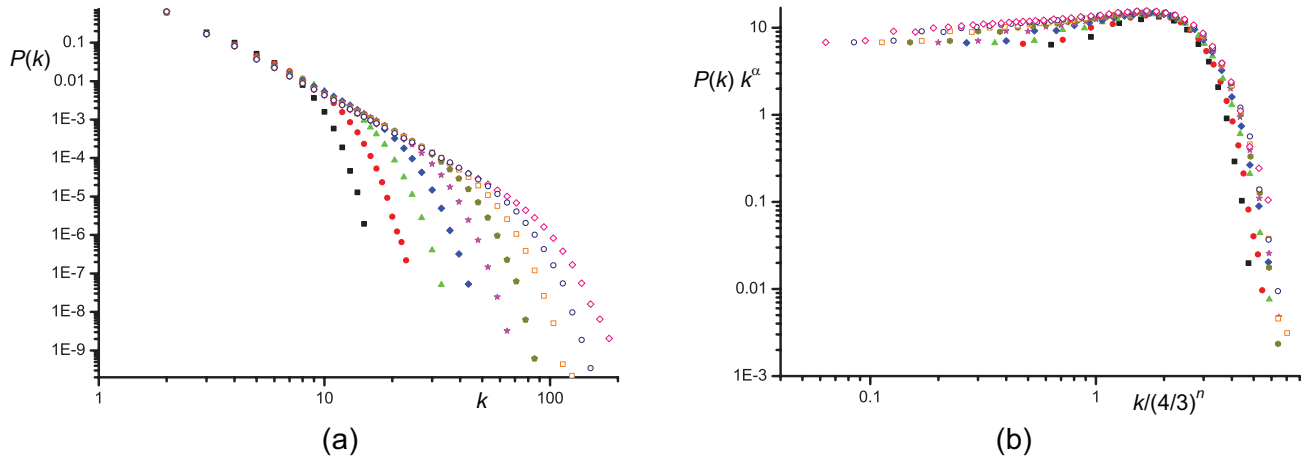, width=17 cm} \caption{(A) Degree distributions of various generations of the tetragon-based networks: generations 4 (black squares), 5 (red circles), 6 (green upward triangles), 7 (blue diamonds), 8 (purple stars), 9 (dark yellow pentagons), 10 (orange open squares), 11 (blue open circles), and 12 (purple open diamonds). The results shown are after averaging over $2\times 10^5$ realizations and logarithmic binning with step 1.1. (B) Same distributions in the rescaled coordinates. The collapse is somewhat worse than in \fig{fig:degree4}(B) presumably because the typical degrees in the hexagon-based network for the same number of nodes.}
\label{fig:degree6}
\end{figure}

\section{Polygon-based networks with smoothly changing exponent $\gamma$}

As a result of the previous section we now have a sequence of Apollonian-like models which generate planar scale-free networks with a discrete sequence of degree-distribution exponents $\alpha_m=1+ \ln 2 / (\ln(m+1)-\ln m),\, m=2,3,\dots$ Is it possible to further generalize the model to make $\alpha$ change continuously, and take any intermediate values, including, for example $\alpha=3$, corresponding to the point where the second moment of the degree distributions diverges for the first time? 

It turns out that this is indeed possible. One way to do that is as follows. Assume that when introducing a new shortcut
dividing a polygon in two we make the resulting partition to be a pair of tetragons with probability $p$ and a pair of hexagons with probability $1-p$. That is to say, if the original polygon is a tetragon then with probability $p$ we
introduce a 2-step path connecting opposite vertices and with probability $q=1-p$ --- a 4-step path, ii) if the
original polygon is a hexagon then with probability $p$ the new path connecting two opposite vertices is a 1-step path with probability $p$, and a 3-step path with probability $q$. We restrict ourselves here to this simplest construction, although it is possible to create more complicated rules. For example, one can introduce correlations between generations in a Markovian way so that there is a matrix $p_{ij}$ of probabilities for a tetragon to give birth to a
couple of tetragons, a hexagon to give birth to a couple of tetragons, etc. As a result it might be possible to construct a network which is, for example, tetragon-dominated at large scales (early generations) and hexagon-dominated at small scales (later  generations). 

Once again, consider a node, which is created at generation $n_0$ and let us study its degree distribution at generation $n+n_0$. This degree distribution depends only on $n$ and on the number of edges of the initial two faces adjacent to the node where tetragons or hexagons. 

Let average fraction of tetragons at a given generation be $p$ and the fraction of hexagons -- $q=1-p$. Then, for each face adjacent to a given node the probability that this face is a tetragon is
\be
\pi (p) =\frac {4p}{4p+6q} = \frac{2p}{3-p}
\label{pi}
\ee
Assume now that different faces adjacent to a node are tetragons (hexagons) independently from each other. Generally speaking, it is not true: when a new edge is created the two faces on the sides of it have a similar number of edges. However, one might expect that as the degree of the node grows the correlations become less and less relevant. In this approximation the probability for a node of degree $k$ to have exactly $l$ adjacent tetragons is
\be
{k \choose {l}} \pi^l (1-\pi)^{k-l}
\ee
When the next generation is created, a new edge adjacent to the node under consideration is created with probability $1/2$ for each tetragon face, and with probability $1/3$ for each hexagon face. Therefore one can write down the following approximate equation for the probability $P(k+r|k,p)$ of the node having degree $k+r$ at the next generation given that it had degree $k$ in the previous one 
\be
P(k+r|k,p) = \sum^k_{l=0}\sum^{r}_{s=0} {k \choose {l}} \pi^l (1-\pi)^{k-l} {l\choose s} \left(\frac{1}{2}\right)^l {k-l \choose r-s} \left(\frac{1}{3}\right)^{r-s}\left(\frac{2}{3}\right)^{k-l-r+s}
\label{prob46}
\ee
where  we assume that bionomial coefficients ${m \choose {n}}$ are zeros if $n>m$ or $n<0$. Now, introduce the probability $F_n(k|p)$ for a node to have degree $k$ $n$ generations after its creation, and the corresponding generation function
\be
f_n(\lambda|p) = \sum^{\infty}_{k=2} F_{n}(k) \lambda ^k.
\ee
Then $f_0 (\lambda|p) = \lambda^2$ and
\be
F_{n}(k) = \sum^{k}_{k'=[\frac{k+1}{2}]} P(k|k',p)F_{n-1}(k')
\ee
and
\be
f_n(\lambda|p) = \sum^{\infty}_{k=2}\sum^{k}_{k'=[\frac{k+1}{2}]} \lambda^k P(k|k',p) F_{n-1}(k') = \sum^{\infty}_{k'=2}F_{n-1}(k') \lambda ^{k'} \sum^{2k'}_{k=k'} \lambda ^{k-k'} P(k|k',p)
\ee
Using \eq{prob46} it is easy to calculate the second sum on the right hand side:
\be
\begin{array}{rll}
\sum^{k'}_{r=0} \lambda ^{r} P(k'+r|k',p) &= &\sum^{k'}_{r=0} \sum^{k'}_{l=0}\sum^{r}_{s=0} \frac{k'!}{s! (l-s)! (r-s)! (k'-l-r+s)!} \pi^l (1-\pi)^{k'-l} \left(\frac{1}{2}\right)^l \left(\frac{1}{3}\right)^{r-s}\left(\frac{2}{3}\right)^{k'-l-r+s} \lambda^r = \medskip \\
& = & \sum^{k'}_{r=0} \sum^{k'}_{l=0}\sum^{r}_{s=0} \frac{k'!}{s! (r-s)! (l-s)! (k'-l-r+s)!} \left(\frac{\pi \lambda}{2}\right)^{s}\left(\frac{\pi}{2}\right)^{l-s}\left(\frac{(1-\pi)\lambda}{3}\right)^{r-s}\left(\frac{2(1-\pi)}{3}\right)^{k'-l-r+s} = \medskip \\
& = & \left(\frac{\pi \lambda}{2} +\frac{\pi}{2} +\frac{(1-\pi)\lambda}{3} +\frac{2(1-\pi)}{3}\right)^{k'}
\end{array}
\ee
which leads to the following equation for the generating function
\be
f_n(\lambda) = f_{n-1}\left(\lambda\left(\frac{4-\pi}{6}+\frac{2+\pi}{6}\lambda\right)\right) = f_{n-1}\left(\frac{2-p+\lambda}{3-p} \lambda \right),
\ee
where we took \eq{pi} into account to get to the last expression. Proceeding as before, we obtain the equations for the generating function of the full limiting degree distribution $g_{\infty}(\lambda)$ (except for the initial set of nodes)
\be
g_{\infty} (\lambda)= \frac{\lambda^2}{2} + \frac{1}{2}g_{\infty}\left(\frac{2-p+\lambda}{3-p} \lambda \right)
\ee
Expanding $g_{\infty} (\lambda)$ for $\lambda = 1 -\epsilon, \; \epsilon \ll 1$ in the form (compare \eq{g_l1_expansion})
\be
\bar g(\lambda) = \sum_{i=0}^{\infty} a_i \epsilon^i + C \Gamma(1-\alpha(p)) \epsilon^{\alpha(p)-1} \sum_{i=0}^{\infty} b_i \epsilon^i,
\label{g_p_expansion}
\ee
and equating the coefficients term by term exactly in the same way as in section II, we get the following equation for the degree distribution exponent $\alpha(p)$: 
\be
2 = \left(\frac{4-p}{3-p}\right)^{\alpha(p)-1} , \;\; \alpha(p)=1+ \frac{\ln 2}{\ln(4-p)-\ln(3-p)}
\label{alpha_p}
\ee
Thus, for example, the interesting case $\alpha(p) =3$ when the second moment of the degree distribution diverges for the first time, corresponds to
\be
p|_{\alpha=3} = 2 -\sqrt{2} \approx 0.58579\dots 
\ee 

\begin{figure}
\epsfig{file=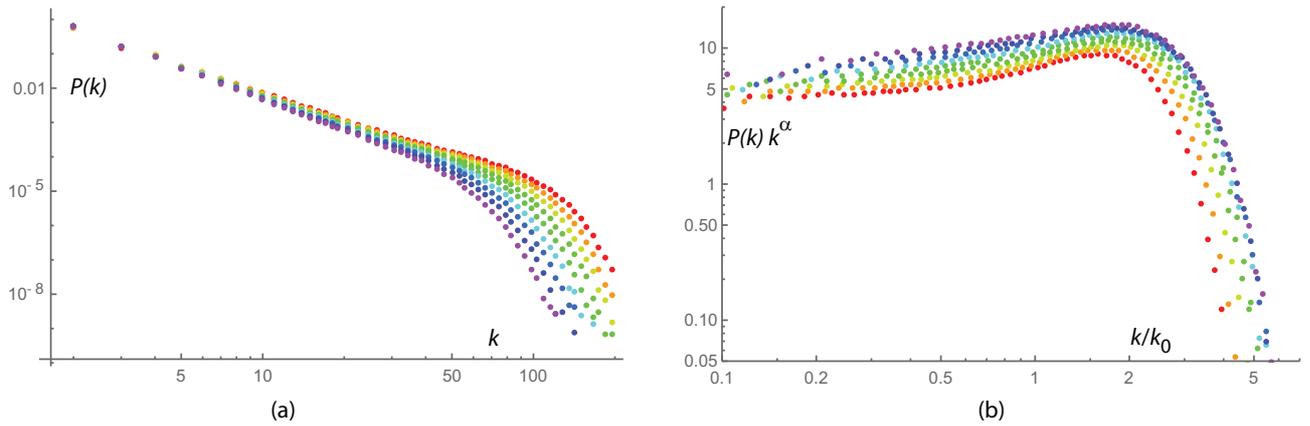, width=17 cm} 
%\centerline{\includegraphics[width=15cm]{apollo_varied_p.pdf}}
\caption{(a) Degree distributions of $n=10$-th generation mixed tetragon-hexagon networks with varying $p$, rainbow changing color from $p=0.9$ (red) to $p=0.1$ (violet) The results shown are after averaging over $10^5$ realizations and logarithmic binning with step 1.05. (B) Same distributions in the rescaled coordinates: $P(k)$ is rescaled by its theoretical behavior $k^{-\alpha}$ with $\alpha$ given by \eq{alpha_p}, $k$ is rescaled by a factor $k_0(p,n) = ((4-p)/(3-p))^n$, which approximates the grows of the maximal accessible degree with the number of generations $n$.}
\label{fig:varied}
\end{figure}

In \fig{fig:varied} the numerical data for the degree distribution of the mixed networks is shown. One can clearly see that the slope of the distribution gradually changes with changing $p$. Moreover, after rescaling the degree distribution with the power law prescribed by \eq{alpha_p} all curves are approximately flat for small $k$, validating the approximation of independent phases.

\section{Discussion}

In this paper we present one possible class of planar random graphs constructed from planar polygones using procedure similar to the construction of Apollonian graphs \cite{apol1}. We show that $2m$-polygone-based graphs have a limiting power law degree distribution with the exponent
\be
\alpha_m = 1+\frac{\ln 2}{\ln(m-1) - \ln m}
\ee
and calculate the moments of the degree distribution. The second moment of the degree distribution diverges as $(9/8)^n$ with the number of generations $n$ in the case of tetragon-based graphs (see \eq{k2all}), and converges to a finite value \eq{moments} for the polygons with larger number of edges. Moreover, as described in the last section, it is possible to construct a mixed model based on two different polygons (tetragons and hexagons in our example) so that on all stages of construction tetragons are formed with probability $p$ and hexagons --- with probability $1-p$. By varying $p$ one can adjust the slope of the degree distribution in order to achieve a desired value in a way reminiscent of evolving Apollonian networks \cite{evolving}.

Clearly, all graph classes presented here are small world. Indeed, the diameter of the graphs grows at most linearly with the number of generations:
\be
d_{n+1}<=d_n + 2 [m/2]
\ee
where $d_n$ is the diameter of the $n$-generation graph, and $[x]$ is the integer part of $x$. In turn, the total number of nodes grows exponentially with the number of generations, thus diameter is at most proportional to the logarithm of the number of nodes. 

The shortest cycles in the graphs presented here are $2m$, and, in particular, there are no triangles in them, so, formally speaking, the clustering coefficient is zero. However, this should not obscure the fact that there is actually a huge number of short cycles in these graphs. Indeed, consider the following auxiliary construction. Let the polygon-based construction be exactly like presented above up to $n$-th generation, but then connect all the nodes belonging to the same face on the last generation of the procedure, so that the smallest faces (i.e., faces constructed on the last step) are considered to be complete graphs $K_{2m}$ ($(2m-1)$-simplices). The large-scale structure of the resulting graph (including, e.g., the slope of the degree distribution) will be exactly the same as in the original polygon-based procedure, but a finite fraction of nodes (those created in the $n$-th generation of the construction) will have clustering coefficient 1, guaranteeing that average clustering coefficient of the whole graph remains finite as $n\to\infty$. In order to use polygon-based graphs as toy model for experimental systems it might be reasonable, instead of adding all possible links connecting the vertices in the smallest faces, add a random fraction of them in order to fit the observed clustering coefficient. 

\begin{figure}
\epsfig{file=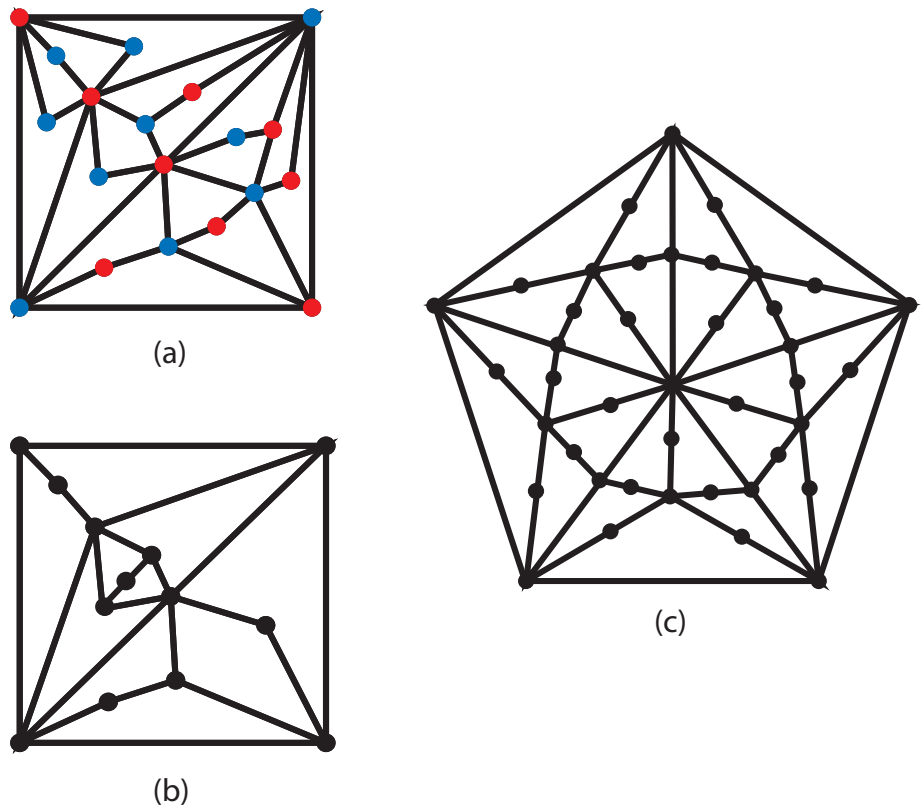, width=12 cm} \caption{(a) Tetragon-based networks are bipartite. (b) An example of a particular realization of a randon tetragon-based graph. (c) an example of a 2-nd generation deterministic pentagon-based graph.}
\label{fig:general}
\end{figure}

Interestingly, polygon-based graphs with even $m$ are bipartite (see \fig{fig:general}(a)). In that sense the tetragon-based graph seems to be a natural generalization of the Apollonian construction for the case of bipartite graphs. We expect that there might be a connection between bipartite polygon-based graphs and space-filling bearings, which allow only cycles of even length \cite{oron} in a way similar to the connection between original Apollonian networks and space-filling systems of embedded disks. Exploring this question goes, however, beyond the scope of this paper.

We present here just one class of possible generalizations of the Apollonian construction based on polygons of arbitrary length. It is quite easy to suggest various other generalizations. The most obvious example is, probably, the random polygon construction, when new graphs are constructed not generation-by-generation by splitting all the polygons of the previous generation at once, but rather by randomly choosing on each step a face to split. In \fig{fig:general}(b) we resent an example realization of such a tetragon-based random graph. Another, this time a completely deterministic construction, is as follows. Consider a polygon with odd number of edges $2m+1, m\geq 1$. Put a point inside the polygon and connect it with all vertices of the polygon by chains of $m$ edges and $(m-1)$ nodes. This splits a polygon into $2m+1$ faces, each having exactly $2m+1$ edges. On the next step, repeat this procedure for each of the faces, and proceed ad infinitum. \fig{fig:general}(c) shows the second generation pentagon-based graph obtained via such procedure. Clearly, this construction is an even more obvious generalization of the Apollonian graph construction (indeed, $m=1$ case is just the Apollonian graph itself). However, it means that it has standard drawbacks of the Apollonian graph in a sense that it is a single deterministic object, rather than a stochastic ensemble of graphs, and that its limiting degree distribution is not a power law but rather a log-periodic function with power law envelope. 

We think that classes of graphs presented here are a useful addition to the toolkit of toy models to model scale-free graphs. Indeed, while having the main advantages of the Apollonian networks, they have additional flexibility in a sense that one might regulate the slope of the degree distribution and the average clustering coefficient in a way described above. In particular, such graphs might be, in our opinion, in the applications where graph planarity is essential \cite{barth}, for example, in quantitative geography in the study of formation of the city systems.

\section*{Acknowledgements}

The authors are grateful to the H. Hermann for encouraging comments on the idea of this work, and to G. Bianconi, S. Nechaev and P. Krapivsky for interesting discussions. MT acknowledges support from the RSF grant 21-11-00215, DK is grateful for the financial support of the BASIS foundation for the support of theoretical physics and mathematics grant 19-2-6-63-1. The research presented here was undertaken in the Laboratory of Nonlinear, Nonequilibrium and Complex Systems at Moscow State University which for the last 10 years has been headed by Prof. M.I. Tribelsky. We are very grateful for his valuable scientific and personal advice, kindness and support over all these years and would like to dedicate this work to him on the occasion of his 70-th birthday.

\end{document}